# Plasmon-Induced Tuning of Cerium Oxidation States in Au@CeO$_x$ Core@Shell Nanoparticles

Klára Beranová[1, a)], Kevin C. Prince[2, 3], Mariana Klementová[4], Marek Vronka[4], Oleksandr Romanyuk[1]

**Affiliations**

[1] FZU - Institute of Physics, Czech Academy of Sciences, Cukrovarnická 10, Prague, 162 00, Czech Republic

[2] Elettra-Sincrotrone Trieste S. C. p. A., in Area Science Park, Strada Statale 14, km 163.5, Basovizza, Trieste, 34149, Italy

[3] Charles University, Faculty of Mathematics and Physics, Department of Surface and Plasma Science, V Holešovičkách 2, Prague, 180 00, Czech Republic

[4] FZU - Institute of Physics, Czech Academy of Sciences, Na Slovance 2, Prague, 182 21, Czech Republic

[a)] corresponding author: klara.beranova@fzu.cz

**ABSTRACT**

CeO$_x$-based nanoforms are widely used in catalysis, or biomedical applications due to their redox activity and oxygen storage capacity. The key parameters determining their surface chemistry are the $Ce^{3+}$/$Ce^{4+}$ ratio and the ability to transition between $Ce^{4+}$ and $Ce^{3+}$ states. We synthesized Au@CeO$_x$ core@shell nanoparticles with different thicknesses of CeO$_x$ shells and different $Ce^{3+}$/$Ce^{4+}$ ratios through a photothermal reaction driven by localized surface plasmon resonances (LSPRs) at the Au nanoparticle surface induced by visible light. We introduce a way to further enhance the $Ce^{3+}$/$Ce^{4+}$ ratio in the shell by exposing the Au@CeO$_x$ nanoparticles to visible light using a green laser (532 nm, 50 mW). Our findings based on photoelectron spectroscopy indicate that the $Ce^{4+}$-to-$Ce^{3+}$ transition results from LSPR-induced superheating of the Au/CeO$_x$ interface, leading to the formation of oxygen vacancies and reduction of $Ce^{4+}$ ions. This process is reversible upon air exposure suggesting that the ability to transition between the $Ce^{4+}$ and $Ce^{3+}$ states is retained in the Au@CeO$_x$ nanoparticles. Our study presents the CeO$_x$-based nanoforms with a tunable cerium valence state ratio, highlighting the potential of plasmonic control in optimizing their photocatalytic and enzyme-mimetic properties.

Cerium oxide, CeO$_x$ ($1.5 < x < 2$) is known for its remarkable oxygen storage capacity, enzyme-mimetic and antioxidant properties.[1–4] These unique properties originate from the ability of cerium to transition between $Ce^{3+}$ and $Ce^{4+}$ states, generating oxygen vacancies in an oxygen deficient environment and scavenging oxygen in an oxygen rich environment. Consequently, CeO$_x$ nanoforms are used in various catalysts or sensors.[5–11] Furthermore, CeO$_x$ nanoforms have been recently investigated for potential use in medical applications, including treatment of cancer or neurodegenerative diseases, due to their ability to reduce oxidative stress by scavenging reactive oxygen species (ROS) and prolonging the lifetime of neural cells.[12–14] On the contrary, some studies reported that CeO$_x$ nanoforms are toxic because of excessive ROS production, or dissolution of CeO$_x$ and release of highly reactive $Ce^{3+}$ ions to the environment.[15,16] The toxicity of CeO$_x$ nanoforms remains ambiguous up to today. The continuous increase of applications using CeO$_x$ nanoforms results in a growing exposure to humans, potentially giving rise to health issues. Therefore, research on fundamental principles and mechanisms determining toxicity of CeO$_x$ nanoforms is essential for safe use of advanced CeO$_x$-based technologies.

The role of cerium valence states in ROS production was proposed by Celardo et al.[3] They concluded that the valence state of cerium is a determining factor for ROS production or scavenging capacity. It was hypothesized that if cerium exists in a greater $Ce^{3+}$/$Ce^{4+}$ ratio, the generation of harmful

ROS will increase and the antioxidant potential of $CeO_x$ will decrease.[2] Further studies demonstrated that the $Ce^{3+}/Ce^{4+}$ ratio is not such a crucial factor in determining the antioxidant potential of $CeO_x$ nanoparticles as is the ability of cerium to transition between the two valence states.[17,18]

The $Ce^{3+}/Ce^{4+}$ ratio in $CeO_x$ depends strongly on the synthesis method, size and shape of the nanoforms.[19,20] Introducing a dopant can modify the $Ce^{3+}/Ce^{4+}$ ratio, and enhance or suppress formation of oxygen vacancies.[17,18,21] Changing the $Ce^{3+}/Ce^{4+}$ ratio in already synthesized $CeO_x$ nanoforms requires application of reagents or an external source of energy, such as UV-light or heat.[22–28] One of the major drawbacks of this approach is possible damage to the surroundings of the nanoparticles. This is especially critical in the case of *in-vitro* and *in-vivo* studies and medical applications.

In this paper, we present a way to tune the $Ce^{3+}/Ce^{4+}$ ratio in $CeO_x$-based nanoforms by combining cerium oxide and gold in an Au@$CeO_x$ core@shell structure. Au nanoparticles were incorporated as cores into the Au@$CeO_x$ system because they are suitable for biomedical studies and applications due to their high biocompatibility and low toxicity.[29] Furthermore, the Au nanoparticles exhibit plasmonic properties, generating localized surface plasmon resonances (LSPRs) at the Au surface by absorbing a portion of visible light, superheating the Au surface and producing surface localized "hot" electrons.[30] LSPRs may induce chemical changes in their closest environments, such as the $CeO_x$ shell, through several different mechanisms described in recent review.[31] However, all these mechanisms ultimately lead to a generation of charge carriers in $CeO_x$ and a short-lived transition between $Ce^{4+}$ and $Ce^{3+}$.[10,11,31–36] In this paper, we introduce a long-lived $Ce^{4+}$ -to- $Ce^{3+}$ reduction, which can be achieved by simple irradiation of Au@$CeO_x$ nanoparticles by visible light. This method exhibits a potential to control and optimize the chemical composition of $CeO_x$ surface with minimal damage to the environment. It is possible to tune the $Ce^{3+}/Ce^{4+}$ ratio in ongoing processes, and without altering size and surface area of the nanoparticles. In this paper, we present the detailed study of the $Ce^{4+}$ -to- $Ce^{3+}$ reduction caused by irradiating the Au@$CeO_x$ nanoparticles by visible light as observed by means of X-ray photoelectron spectroscopy (XPS).

The whole synthesis protocol is presented in the Supplementary Material. To summarize, the Au cores were prepared following the protocol published by Acres et al.[37] As a result, we obtained a ruby red colloid of approximately spherical citrate-stabilized Au nanoparticles with average diameter of 17 nm (see Figure 1(a) and 1(b)). Coating the Au nanoparticles by the $CeO_x$ shell was achieved by employing LSPRs photothermal induced chemical reactions similarly to the report of Zhong et al.[38] $Ce(NO_3)_3{\cdot}6H_2O$, citric acid and ethylene glycol were mixed in deionized water. After adjusting pH as close to 7.0 as possible, the dialyzed Au colloid was added, and the solution was mixed in the dark. A resulting purplish solution was irradiated by visible light. The temperature of the solution was kept at room temperature. Because the Ce-gel is formed only above 80°C,[38] the irradiation of the solution by visible light and the LSPR-generated surface heat induced growth of the Ce-gel solely on the surface of the Au nanoparticles. Using different ratios between the Au- and Ce-based solutions, irradiation periods, and sources of visible light, such as a Xe-lamp, green LED or sunlight, resulted in synthesizing a set of Au@$CeO_x$ samples with the same size of cores and different thicknesses of the shell. After the irradiation, the solutions containing the Au@$CeO_x$ nanoparticles were dialyzed to remove any unreacted reagents. The resulting solutions were dried at slightly elevated temperature. The sediment was then calcined at 400°C in air to transform the Ce-gel to $CeO_x$. The samples were labelled as Ce with added number corresponding to an average thickness of $CeO_x$ shells.

The structure and spatial composition of the Au@$CeO_x$ nanoparticles were investigated by transmission electron microscopy (TEM) and scanning transmission electron microscopy (STEM). A 200 kV- field emission transmission microscope FEI Tecnai X-Twin F20 equipped with energy-dispersive spectrometry (EDX) was used for analysis. Figure 1(c) shows a High Angle Annular Dark Field (HAADF) micrograph of the Ce16.5 sample along with EDX maps of the selected NP. The HAADF micrograph enables distinguishing the sample parts consisting of light elements from the parts containing heavy elements due to incoherent scattering on the atoms of the sample. The EDX maps show clear distinction between the Au core and the Ce-O composed shell, thus confirming the core@shell structure of our Au@$CeO_x$ nanoparticles. The thicknesses of the shells were not entirely uniform, and particles exhibited severe agglomeration. However, all observed nanoforms exhibited the core@shell structure. Due to these effects, it was difficult to determine the average thickness of the $CeO_x$ shells from the micrographs. Therefore, we chose to estimate an average $CeO_x$ thickness via simulating the photoelectron spectra, which provide macroscopic information by averaging over large number of nanoparticles.

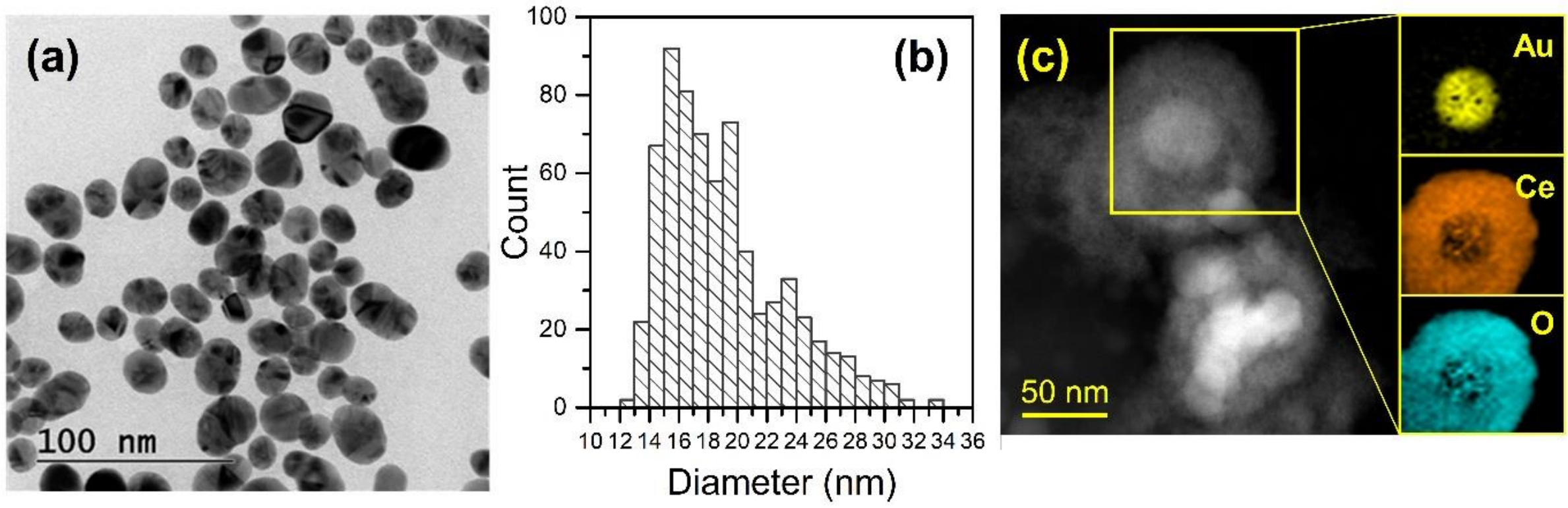

*Figure 1: (a) TEM micrograph of Au NPs, (b) distribution of Au NPs sized taken from several micrographs, (c) HAADF micrograph of an Au@$CeO_x$ NP from the sample Ce16.5 with EDX elemental maps of gold (yellow), cerium (orange) and oxygen (blue).*

XPS was used to determine elemental composition of the samples, bonding states of elements and valency of cerium ions. XPS measurements were performed using the AXIS Supra photoelectron spectrometer (Kratos Analytical). The Au@$CeO_x$ nanoparticles were mixed with deionized water and dropped on pieces of silicon wafer, which were attached by an insulating paper tape to a sample holder. Monochromatized Al Kα along with an electron charge-compensation flood gun were used for all measurements. High resolution XPS spectra of Ce 3d, O 1s, Au 4f, C 1s, and Si 2p/Si 2s regions were acquired. The total intensities of these spectra were used as a benchmark for the output of the SESSA simulations.

The graphical user interface (GUI) of the Simulation of Electron Spectra for Surface Analysis (SESSA) - Version 2.1[39] software was used to model the Au@$CeO_x$ nanoparticles and generate their photoelectron spectra. The model of the sample was set as 4-layered spheres: $H_2O$/C/$CeO_x$/Au (Au is the innermost layer) equidistantly spread over a $SiO_2$ substrate. The geometrical parameters of the model were varied to achieve the best agreement with photoelectron spectra acquired for the Au@$CeO_x$ samples. More details regarding the SESSA simulations, including the geometrical parameters of the best-matching models, can be found in the Supplementary Material. In all simulated Au@$CeO_x$ samples, the Au core size remained constant at 18.50 nm. The samples were designated as Ce0.6, Ce0.9, Ce1.2, Ce1.4, Ce1.8, Ce2.3, Ce8.5, Ce13.7, Ce16.0 and Ce16.5, according to the simulated average $CeO_x$ thickness. The simulated size of the Au cores shows a very good agreement with our TEM results (see Figure 1(a) and 1(b)). The simulated thickness of the sample Ce16.5 shell is also close to that of the nanoparticle presented in Figure 1(c) supporting reliability of the simulations.

The valence state of cerium in the shells was determined by deconvolution of the Ce 3d photoelectron spectra using the KolXPD – Version 1.8.0 (build 68) software. The background used for the Ce 3d region was set as a combination of three ranged linear backgrounds defining a general outline of the broad region, which was superimposed with a Shirley background. The Ce 3d spectra were fitted by five doublets of Voigt profile, which represent different Ce 4f final state configurations.[40] Three $3d_{3/2}$ – $3d_{5/2}$ doublets correspond to the $Ce^{4+}$ oxidation state ($f^0$: 916.7 – 898.3 eV, $f^1$: 907.1 – 888.7 eV, and $f^2$: 900.6 – 882.2 eV) and two doublets to the $Ce^{3+}$ oxidation state ($f^1$: 899.5 – 881.1 eV, $f^0$: 903.8 – 885.4 eV). It should be noted that the $Ce^{4+}$ $f^2$ doublet is asymmetric, and thus was fitted by a sum of two symmetric doublets.[21,41] Furthermore, $Ce^{4+}$ $f^1$ was fitted by two separate peaks, because their widths significantly differed. The average oxidation state of cerium in the Au@$CeO_x$ nanoparticles was expressed by a parameter *x*, which represents the ratio between cerium of mixed $Ce^{3+}$ - $Ce^{4+}$ valency and oxygen in $CeO_x$ ranging from 1.5 ($Ce_2O_3$ - all cerium ions are $Ce^{3+}$) to 2 ($CeO_2$ - all cerium ions are $Ce^{4+}$). *X* was calculated from deconvolution of the Ce 3d core-level spectra[42] according to the formula

$$x = 2 - \frac{1}{2} \cdot \frac{A(Ce^{3+}\,f^1) + A(Ce^{3+}\,f^2)}{A(Ce^{4+}\,f^0) + A(Ce^{4+}\,f^1) + A(Ce^{4+}\,f^2) + A(Ce^{3+}\,f^2) + A(Ce^{3+}\,f^1)}, \qquad (1)$$

where A denotes the area of a corresponding Ce 3d doublet.

An example of the Ce 3d fit is plotted in Figure 2(a) for the sample Ce16.5. Values of *x* determined via equation (1) of the initial states of the Au@$CeO_x$ samples (*As received*) are plotted in Figure 2(b) as a function of the simulated average $CeO_x$ shell thickness. We can see that *x* in $CeO_x$ strongly depends on the shell thickness, especially in the case of the sub-nanometer shells. Figure 2(b) also shows a simulated average thickness of water overlayer, which exhibits an inverse tendency to *x* in $CeO_x$.

The correlation between adsorbed water and $CeO_x$ thickness may be explained by a strong affinity of $OH^-$ to $Ce^{3+}$ ions.[43,44] Van Dao reported that the Au-$CeO_2$ interaction increases the $Ce^{3+}$ concentration in $CeO_x$.[35] Dvořák et al showed that $H_2O$ directly reacts with $CeO_x$ leading to incorporation of $OH^-$ groups into the oxide bulk.[8] Rodriguez et al demonstrated that the Au-$CeO_2$ interface is an extremely active place for $H_2O$ dissociation, resulting in formation of oxygen vacancies in $CeO_x$ and reduction of $Ce^{4+}$ to $Ce^{3+}$.[45] Therefore, the large amount of water species simulated at the samples with sub-nanometer thin $CeO_x$ shells and high concentration of $Ce^{3+}$ ions can be explained by effective incorporation of water species ($H_2O$ and $OH^-$) into the bulk of the oxide shell. This effect is suppressed for thicker $CeO_x$ shells because the Au-interface is not so exposed to the moisture, which we deem to be the origin of the water species.

The different Au@$CeO_x$ samples consisted of $CeO_x$ shells with different thicknesses and $Ce^{3+}/Ce^{4+}$ ratios. Figure 2(a) shows that the $Ce^{3+}/Ce^{4+}$ ratio in the shell can be further modified by exposing the samples in a vacuum chamber to visible light irradiation. For this purpose, we used a green laser (50 mW, 532 nm) with a wavelength corresponding to the maximum absorbance of Au NPs of this size.[37] Au NPs act as a visible light absorber and LSPRs generator. The size and shape of the NPs determines which part of visible spectrum will be absorbed. Adding cerium oxide does not have much influence on the absorbance since its absorbance is in the UV range.[10,36]

The samples were irradiated by the laser *in-situ* in multiple steps. After every laser irradiation, XPS measurements were conducted in dark to determine the $Ce^{3+}/Ce^{4+}$ ratio. The Ce 3d spectra of the samples Ce0.6, Ce0.9 and Ce16.5 in their initial state and after the last step of laser irradiation are plotted in Figure 2(a). These samples were selected to show three distinctly different $CeO_x$, one with prevalent $Ce^{4+}$ states (Ce16.5), the other with the highest $Ce^{3+}$ concentration (Ce0.6) and the last with approximately same $Ce^{3+}$ and $Ce^{4+}$ concentrations (Ce0.9). An increase in the $Ce^{3+}$ concentration following laser irradiation is apparent from the Ce 3d spectra. This effect is more clearly illustrated in Figure 2(c), which depicts the evolution of *x* in $CeO_x$ with increased time of laser irradiation. A decrease in *x* was observed for all examined samples (shown in Supplementary Materials). The extent of $Ce^{4+}$ - to- $Ce^{3+}$ reduction seems to be influenced by the shell thickness and its initial $Ce^{3+}/Ce^{4+}$ ratio.

Further experiments with the sample Ce0.9 confirmed that the $Ce^{3+}/Ce^{4+}$ ratio in the $CeO_x$ shell remains stable in vacuum conditions for an extended period and that the $CeO_x$ shell can restore its initial $Ce^{3+}/Ce^{4+}$ ratio upon air exposure. After irradiating the sample with the laser in UHV for a total of 60 min, followed by XPS measurements, the sample Ce0.9 was removed from the vacuum chamber and left in air. It was then reinserted and subjected to another 60 min of laser irradiation. *X* estimated from measurements before air exposure (pink circles in Figure 2(c)) closely match those from the cycle after air exposure (gray circles in Figure 2(c)) and remain unchanged even after 90 min of resting in vacuum (the full gray circle). This indicates that $Ce^{3+}$ states formed during laser irradiation in vacuum have long lifetime. Furthermore, the $Ce^{4+}$ -to- $Ce^{3+}$ transition is reversible, which is crucial for catalytic activity and efficiency of surface chemistry processes.

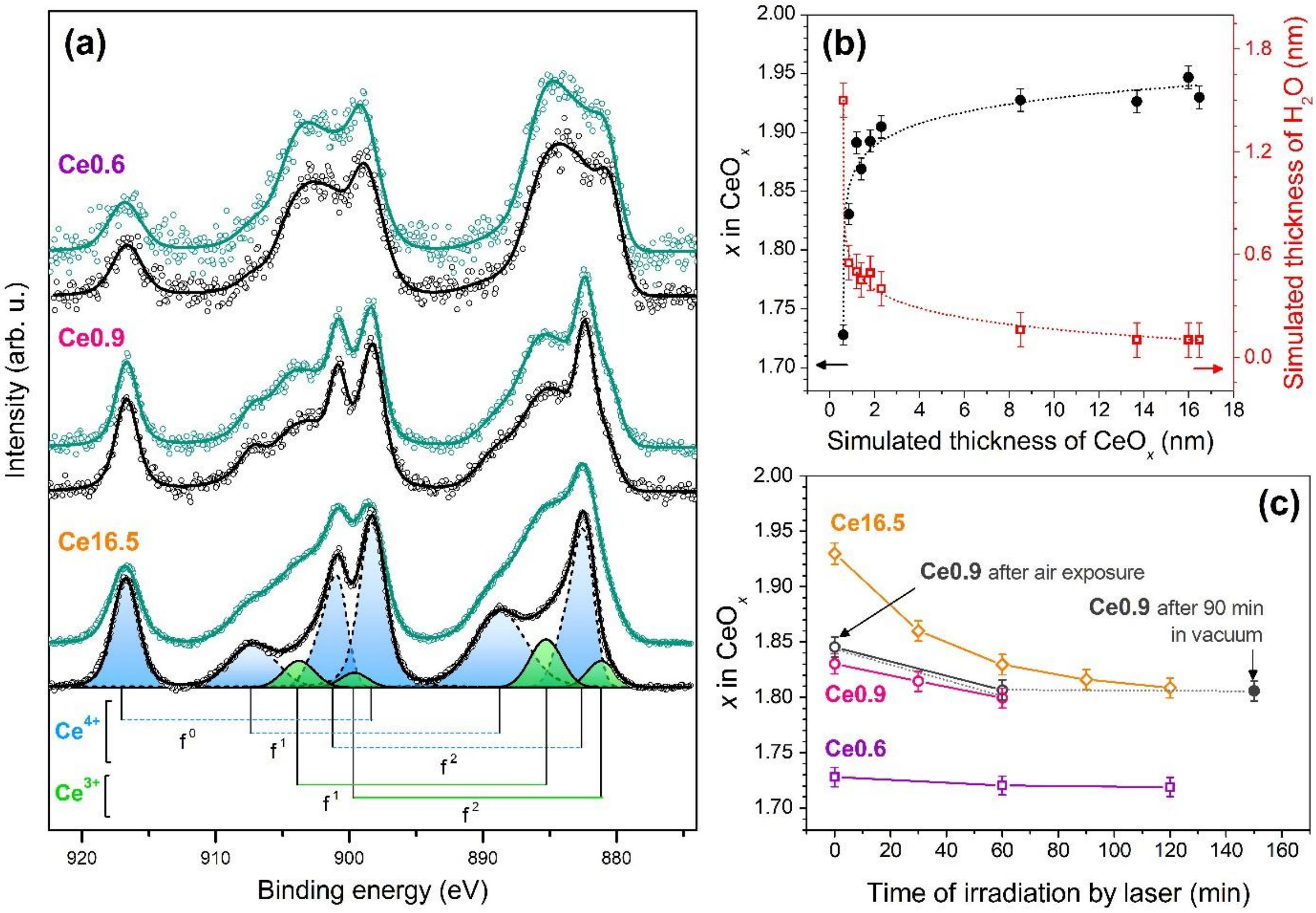

*Figure 2: (a) XPS Ce 3d region of Ce0.6, Ce0.9 and Ce16.5 samples acquired in their initial state (black) and after all irradiating by laser (cyan). Circles represent measured data, and full lines fit results. Spectra are presented normalized to maximum and with subtracted backgrounds for better clarity. $Ce^{3+}$ (green) and $Ce^{4+}$ (blue) fitting components are presented for the sample Ce16.5 - As received. (b) Dependences of x in $CeO_x$ derived from fitting the Ce 3d spectra (left) and simulated average thickness of water overlayer (right) on the simulated average thickness of the $CeO_x$ shell. Dotted lines are plotted for guiding the eyes. (c) Evolution of x in $CeO_x$ with increasing time of irradiation by laser for samples Ce0.6 (purple squares), Ce0.9 (pink circles), and Ce16.5 (orange rectangles). The gray circles represent the sample Ce0.9 after air exposure. The full gray circle at 150 min was obtained after 90 min of resting in UHV. The steps are line-connected for guiding the eyes and indicating chronology of measurements.*

It is evident that the long-lived $Ce^{4+}$ -to- $Ce^{3+}$ transition in the $CeO_x$ shells is driven by LSPRs at the Au core surface. Laser irradiation of pure $CeO_x$, or $CeO_x$ mixed with Au NPs did not result in any changes in the Ce 3d spectra (data not shown). In the Au@$CeO_x$ core@shell system, the $Ce^{4+}$ -to- $Ce^{3+}$ transition can occur through two mechanisms. The first one involves accommodation of an electron from a donor, such as Au core.[10,11,28] However, the lifetime of charge carriers in $CeO_x$ is rather short because of facile migration and fast recombination.[34,35] The second mechanism involves formation of oxygen vacancies and the desorption of oxygen.[2,42]

Although we did not observe any increase in pressure within the vacuum chamber during the laser irradiation, a gradual depletion of oxygen from $CeO_x$ is evident from the atomic concentrations of oxygen species bound to cerium ions, $O_{Ce}$, as detailed in the Supplementary Material. The $O_{Ce}$ includes oxygen from the $CeO_2$ lattice, as well as water species bound to $CeO_x$ near the oxygen vacancy sites, such as $OH^-$ and $H_2O$. The gradual decrease in $O_{Ce}$ following laser irradiation aligns with decrease of $x$ in $CeO_x$. This suggests that oxygen desorption occurs and that the long-lived $Ce^{4+}$ -to- $Ce^{3+}$ transition is caused by the oxygen vacancy formation, probably photothermally driven. Our hypothesis is supported by DFT calculations predicting significant lowering of oxygen vacancy formation energy at the Au@$CeO_2$ core@shell NPs.[35]

In summary, we have demonstrated a method for synthesizing Au@$CeO_x$ core@shell nanoparticles with varying $CeO_x$ shell thicknesses using LSPR photothermal-induced coating. The $Ce^{3+}/Ce^{4+}$ ratio depends on the thickness of the shell and can be further increased by exposing the samples to visible light under vacuum. We suggested that the $Ce^{4+}$ -to- $Ce^{3+}$ transition occurring in the $CeO_x$ shell upon visible light irradiation is caused by oxygen vacancy formation driven by LSPRs-induced superheating of the Au core surface. Additionally, we demonstrated that the $Ce^{3+}/Ce^{4+}$ ratio remains stable for extended period in vacuum conditions and that the $Ce^{4+}$ -to- $Ce^{3+}$ transition is reversible, suggesting that redox cycling in the Au@$CeO_x$ nanoparticles is not suppressed.

The possibility to tune the $Ce^{3+}/Ce^{4+}$ ratio in the Au@$CeO_x$ nanoparticles without any use of additive chemicals or damaging UV-light has tremendous potential for applications in medical nanoscience or photocatalysis. Therefore, this method can be employed in various applications which require tuning the $CeO_x$ physiochemical properties without damaging the environment. Additionally, this tuning can be area-specific, which is essential in targeted treatments. Although the tuneability of the $Ce^{3+}/Ce^{4+}$ ratio is incredibly promising, it is necessary to determine its extent in oxidizing environments, such as atmospheric, aquatic or intra cellular environments. However, this is a topic for further research.

See the Supplementary Material for additional data and analysis that supports the findings of this study. Section 1 contains detailed information about used materials and the synthesis protocol. Section 2 consists of details of simulations in SESSA. Section 3 contains details concerning measurement and analysis of the XPS spectra.

This work was supported by the Czech Academy of Sciences—Strategy AV21 and by the Ministry of Education, Youth and Sports of the Czech Republic (CzechNanoLab Research Infrastructure, MEYS CR project n. LM2023051).

The authors have no conflicts to disclose.

Klára Beranová: Writing - original draft, Investigation, Formal analysis, Methodology, Visualization; Kevin. C. Prince: Conceptualization, Writing – review; Mariana Klementová: Investigation; Marek Vronka: Investigation, Writing – review; Oleksandr Romanyuk: Writing – review, Funding acquisition

The data that support the findings of this study are available from the corresponding author upon reasonable request.

# Supplementary Material

## Plasmon-Induced Tuning of Cerium Oxidation States in Au@CeO$_x$ Core@Shell Nanoparticles

Klára Beranová[1, a)], Kevin C. Prince[2, 3], Mariana Klementová[4], Marek Vronka[4], Oleksandr Romanyuk[1]

**Affiliations**

[1] FZU - Institute of Physics, Czech Academy of Sciences, Cukrovarnická 10, Prague, 162 00, Czech Republic

[2] Elettra-Sincrotrone Trieste S. C. p. A., in Area Science Park, Strada Statale 14, km 163.5, Basovizza, Trieste, 34149, Italy

[3] Charles University, Faculty of Mathematics and Physics, Department of Surface and Plasma Science, V Holešovičkách 2, Prague, 180 00, Czech Republic

[4] FZU - Institute of Physics, Czech Academy of Sciences, Na Slovance 2, Prague, 182 21, Czech Republic

[a)] corresponding author: klara.beranova@fzu.cz

## 1. Synthesis of Au@CeO$_x$ nanoparticles

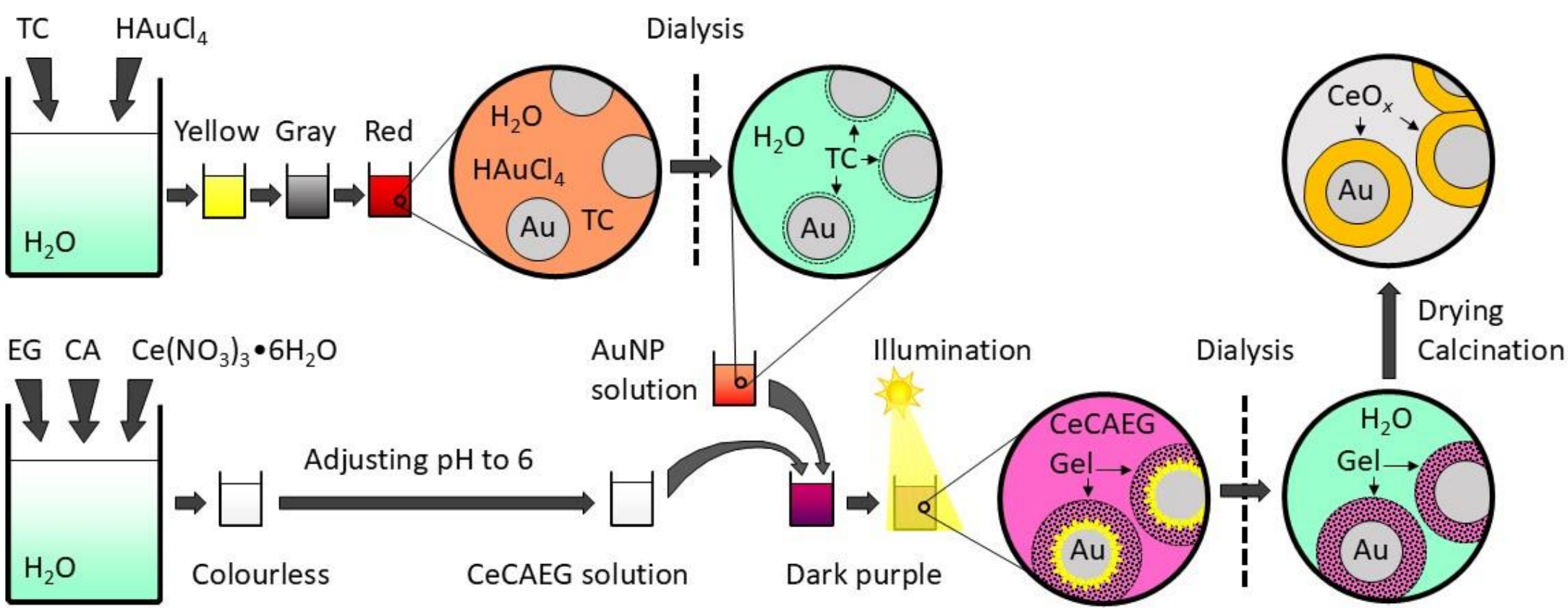

*Figure S1: Schematic illustration of synthesizing the Au@CeO$_x$ NPs.*

### 1.1 Materials

$AuCl_3$: gold(III) chloride (99.99%, Merck Life Science); HCl: hydrochloric acid (36 % w/w aqueous solution, Alfa Aesar); TC: trisodium citrate dihydrate (99 %, Alfa Aesar); $Ce(NO_3)_3 \cdot 6H_2O$: cerium(III) nitrate hexahydrate (99.99 %, REO, Alfa Aesar); EG: ethylene glycol (99 %, Alfa Aesar); CA: anhydrous citric acid (99.5+ %, ACS, Alfa Aesar); $NH_4OH$: ammonia (28.0 – 30.0 % $NH_3$ aqueous solution, ACS, Alfa Aesar).

## 1.2 Synthesis of Au cores

The whole synthesis procedure is shown schematically in Figure S1. Au nanoparticles (AuNPs) were prepared according to the procedure described by Acres et al.[1] $AuCl_3$ was dissolved in 0.5 % (w/v) HCl producing a 4 % (w/v) $HAuCl_4$ solution. 0.2 ml of 4 % (w/v) $HAuCl_4$ was added to 80 ml of ultra-pure water. The solution was slowly heated to just below the boiling point while stirring by a magnetic stirrer. Then, 1 ml of the 1 % (w/v) TC solution was added dropwise into the heated aqueous $HAuCl_4$ solution. The solution slowly changed color from yellow to gray, then to black and finally, it turned red. After reaching ruby red color, it was boiled for 10 min without stirring. The solution was left to cool down, then filtered by a 20 µm syringe filter and then dialyzed by a thoroughly washed cellulose ester dialysis membrane (8 – 10 kD, Spectrum Laboratories) to remove any residual reagents. Deionized water was changed four times in a span of 12 hours. As a result, we obtained round, approximately spherical Au nanoparticles with an average diameter of 17 nm.

## 1.3 Coating by $CeO_x$

Coating by cerium oxide was performed using the LSPRs photothermal induced interface reaction proposed by Zhong et al. for Ag@$CeO_2$ NPs.[2] 174 µg of Ce, 126 µg of CA and 135 µl of EG were added to 100 ml of deionized water and stirred for 1 hour. After that, the pH of the solution was adjusted as close to 6.7 as possible by adding 0.001 M ammonia, producing a transparent colorless CeCAEG solution. The CeCAEG solution was mixed with the AuNP solution and mixed for at least 15 min in the dark. The solution became dark purple. After this step, it was stirred and irradiated by visible light (green LED lamp, Xe lamp, or sunlight). A CeCAEG sol-gel formed solely at the noble metal core due to NPs surface localized heating induced by absorption of visible light. The temperature was checked and controlled to not exceed 25°C. The resulting solutions were dialyzed in the dark to remove any residual reagents, while changing the water four times in a span of 12 hours. The dialyzed solution was dried at a slightly elevated temperature. The sediment was calcined at 400°C in air on glass or Si substrates.

We prepared four different samples while varying the time of irradiation during the exposure, a light source, or the ratio between the Au NPs and CeCAEG solutions. The parameters of sample preparation and their designations are listed in Table S-I.

*Table S-I: Preparation parameters of Au@$CeO_x$ NPs, simulated thicknesses of $CeO_x$ shells, and initial oxidation state of cerium oxide derived from fitting the Ce 3d XPS region (x in $CeO_x$)*

| Sample | Irradiation time (min) | Light source | pH of CeCAEG | AuNP/CeCAEG (v/v) ratio | $CeO_x$ simulated thickness (nm) | $X$ in $CeO_x$ initial oxidation state |
|---|---|---|---|---|---|---|
| Ce0.6 | 15 | Green LED | 6.3 | 1:1 | 0.60 | 1.73 |
| Ce0.9 | 120 | Xe | 6.7 | 1:1 | 0.85 | 1.83 |
| Ce1.2 | 120 | Xe | 6.7 | 1:1 | 1.20 | 1.89 |
| Ce1.4 | 30 | Xe | 6.7 | 1:1 | 1.40 | 1.87 |
| Ce1.8 | 60 | Xe | 6.7 | 1:1 | 1.80 | 1.89 |
| Ce2.3 | 15 | Xe | 6.7 | 1:1 | 2.30 | 1.90 |
| Ce8.5 | 120 | Xe | 6.8 | 1:2 | 8.50 | 1.93 |
| Ce13.7 | 30 | Xe | 6.4 | 1:2 | 13.7 | 1.93 |
| Ce16.0 | 60 | Sunlight | 6.4 | 1:2 | 16.00 | 1.95 |
| Ce16.5 | 15 | Xe | 6.4 | 1:2 | 16.50 | 1.93 |

## 2. Sessa simulations

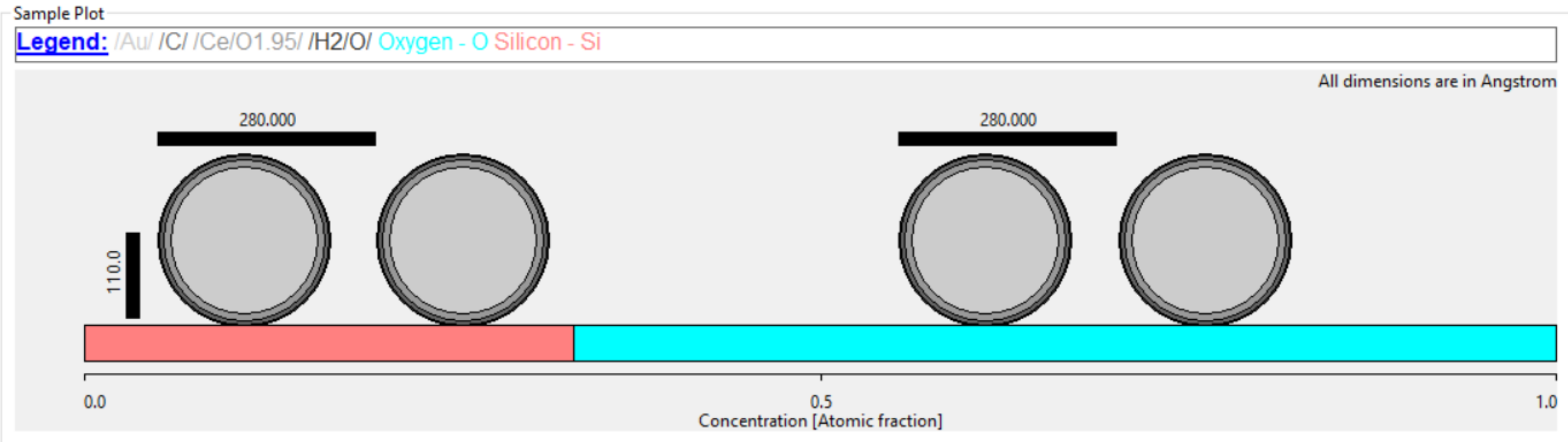

*Figure S2: Illustration of the geometrical set-up of Au@$CeO_x$ NPs (sample Ce16.0) in SESSA.*

The dimensions of the Au@$CeO_x$ nanoparticles were roughly estimated using the Simulation of Electron Spectra for Surface Analysis (SESSA)- Version 2.1 software using the GUI interface. The sample model was set as 4-layered spheres spread over a $SiO_2$ substrate, as demonstrated in Figure S2. The 4$^{th}$ innermost layer, the core, was set as Au spheres. The 3$^{rd}$ layer was set as the $CeO_x$ shell. The ratio between oxygen and cerium in the $CeO_x$ shell was set according to results from analysis of the Ce 3d XPS *As received* spectra. The 2$^{nd}$ was a layer of carbon contaminants and the 1$^{st}$, the topmost layer, consisted of molecular water. The variable parameters were thicknesses of all four layers and distances between the layered spheres. The parameters were changed manually and the simulated intensities of Ce 3d, O 1s, Au 4f, C 1s and Si 2p/ Si 2s regions were compared with intensities acquired by XPS to achieve best agreement between simulations and data. The spectral lines were set as gaussian peaks. In case of more complicated structure, such as Ce 3d, the line shape was set to match the acquired XPS spectrum.

The simulation is compared with the photoelectron spectra for the sample Ce0.9 in Figure S3.

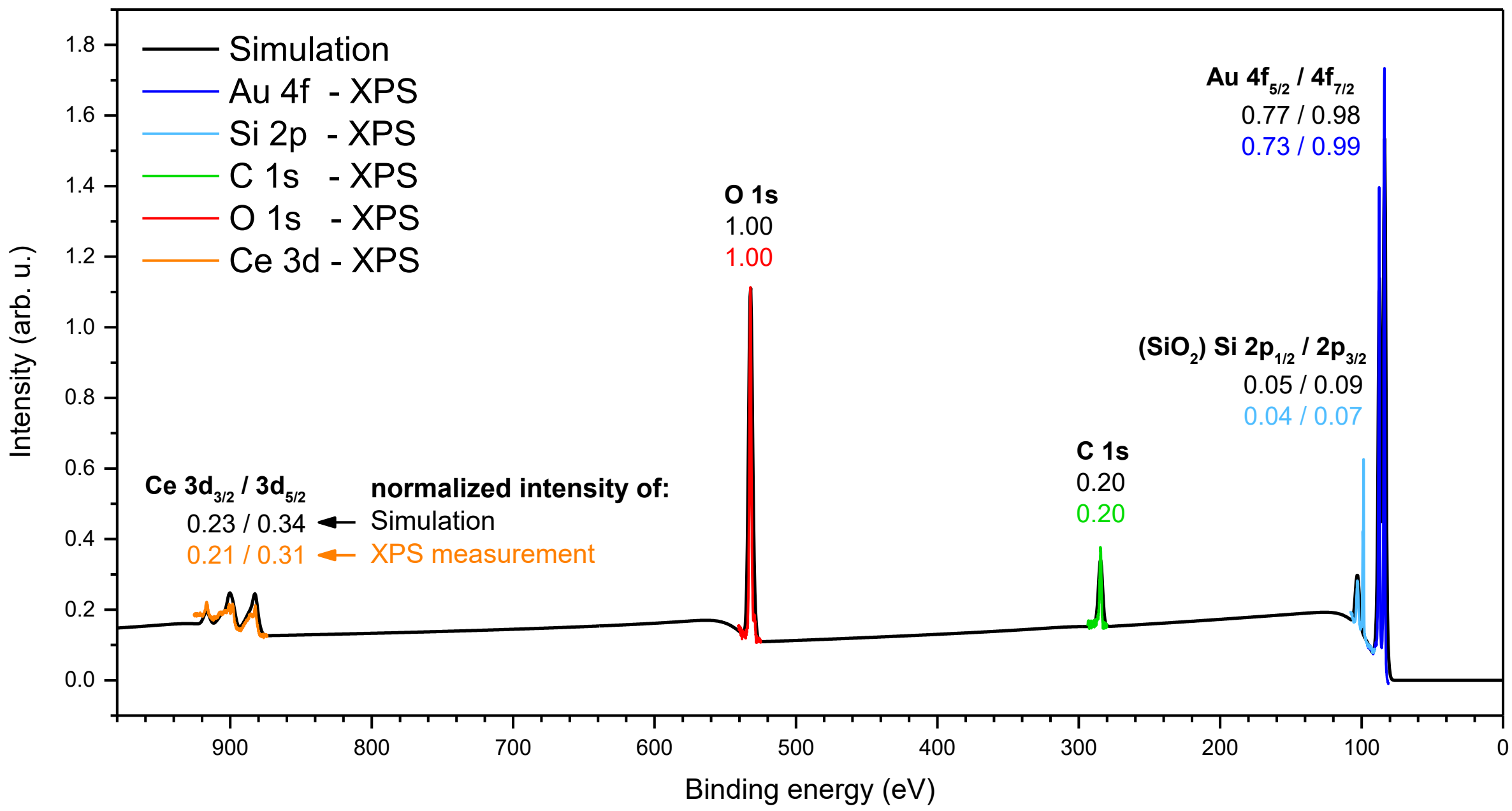

*Figure S3: Comparison of the optimal simulation and XPS experimental data for the sample Ce0.9. The XPS data and the simulation result were normalized to the O 1s intensity.*

The non-variable parameters used in the simulations, such as band gaps, densities, and inelastic mean free paths are listed in Table S-II.

*Table S-II: The list of the optimal layer thicknesses, radii of layered spheres and distances between centers of spheres for all simulated samples, and non-variable input parameters in the SESSA simulations for all layers (band gaps, densities, inelastic mean free paths).*

| Sample | Thickness (nm) | | | | Radius (nm) | Distance (nm) |
|---|---|---|---|---|---|---|
| | Au | $CeO_x$ | C | $H_2O$ | | |
| Ce0.6 | 18.50 | 0.60 | 0.50 | 1.50 | 10.55 | 90.00 |
| Ce0.9 | 18.50 | 0.85 | 1.10 | 0.55 | 10.50 | 26.20 |
| Ce1.2 | 18.50 | 1.20 | 0.80 | 0.50 | 10.50 | 24.70 |
| Ce1.4 | 18.50 | 1.40 | 0.40 | 0.45 | 10.38 | 25.00 |
| Ce1.8 | 18.50 | 1.80 | 0.90 | 0.49 | 10.85 | 28.00 |
| Ce2.3 | 18.50 | 2.30 | 0.60 | 0.40 | 10.90 | 23.00 |
| Ce8.5 | 18.50 | 8.50 | 0.85 | 0.16 | 14.05 | 74.00 |
| Ce13.7 | 18.50 | 13.70 | 0.16 | 0.10 | 16.23 | 34.29 |
| Ce16.0 | 18.50 | 16.00 | 0.15 | 0.10 | 17.38 | 35.00 |
| Ce16.5 | 18.50 | 16.49 | 0.16 | 0.10 | 17.63 | 35.25 |

| Material | Band gap (eV) | | Density x $10^{22}$ (atoms/$cm^3$) [a] | | | |
|---|---|---|---|---|---|---|
| | | | $CeO_{1.73}$ | $CeO_{1.83}$ | $CeO_{1.93}$ | $CeO_{1.95}$ |
| $SiO_2$ | 8.5 [3] | 6.585 | | | | |
| Au | 0.0 | 5.901 | | | | |
| $CeO_x$ | 3.2 [4] | | 4.228 | 4.256 | 4.281 | 4.286 |
| C | 2.8 [5] | 11.030 | | | | |
| $H_2O$ | 6.9 [6] | 10.030 | | | | |

| | Inelastic mean free path (nm) [b] | | | | | | | | |
|---|---|---|---|---|---|---|---|---|---|
| | Au 4f | | Ce 3d | | C 1s | O 1s | Si 2p [c] | | Si 2s |
| | 5/2 | 7/2 | 3/2 | 5/2 | | | 1/2 | 3/2 | |
| $SiO_2$ | 3.835 | 3.842 | 1.962 | 2.007 | 3.403 | 2.852 | 3.800 | 3.802 | 3.692 |
| Au | 1.571 | 1.574 | 0.841 | 0.858 | 1.404 | 1.189 | 1.558 | 1.558 | 1.516 |
| $CeO_x$ | 2.954 | 2.960 | 1.513 | 1.548 | 2.623 | 2.200 | 2.924 | 2.925 | 2.845 |
| C | 3.525 | 3.532 | 1.806 | 1.847 | 3.129 | 2.623 | 3.493 | 3.495 | 3.394 |
| $H_2O$ | 4.489 | 4.499 | 2.262 | 2.315 | 3.975 | 3.319 | 4.484 | 4.450 | 4.319 |

a) Provided by SESSA software
b) IMFP database: TPP-2M formula, 50 eV – 3 keV [IMFP01], sample Ce16.0
c) Sample Ce0.9

# 3. Photoelectron spectroscopy (XPS)

## 3.1 Experimental details

XPS was used to determine elemental composition of the samples, as well as bonding and valence states of elements. XPS measurements were performed using the AXIS Supra photoelectron spectrometer (Kratos Analytical), which enables collecting XPS signal from a 0.8 $mm^2$ small spot. The nanoparticles supported by pieces of Si wafer (N-type doped by phosphorus, 1.35 – 1.40 Ωcm, with (111) surface orientation, Terosil) were attached to the sample holder by an insulating paper tape. Monochromatized Al Kα with emission of 10 mA along with an electron flood-gun charge-compensation were used for all measurements. The survey spectra were collected from 1200 to -5 eV to identify any elements. Additionally, high resolution spectra of Ce 3d, O 1s, Au 4f, C 1s, N 1s and Si 2p/Si 2s regions were acquired for purposes of more detailed analysis (with high signal from cerium, the Ce 4d region overlapped with the Si 2p region therefore, we measured the Si 2s region instead).

The 50 mW-green laser (532 nm) installed outside the UHV chamber was used to irradiate the samples. Notably, the irradiation was performed *in-situ*, in vacuum conditions ($5 \cdot 10^{-9}$ Torr). The area of XPS signal collection was smaller than the area irradiated by the laser, as demonstrated in Figure S4. The laser spot was aligned with the measurement area using images from an optical microscope integrated in the XPS spectrometer. To ensure that the XPS acquisition area was smaller than the irradiated area, the laser beam was slightly defocused. After characterizing the samples in the "As received" state, the selected samples were irradiated by the laser and analyzed by XPS in following cycles:

- Irradiating the sample by the green laser for a specific time (usually 30 min)
- Acquiring the XPS spectra in dark (the laser and the in-situ lamp were switched off during the whole XPS measurement).

The evolution of $x$ in $CeO_x$ with prolonged laser irradiation is plotted in Figure S5 (a) for samples Ce16.5, Ce16.0, Ce13.7, Ce1.8, Ce0.9, Ce0.6.

To minimize the distortion of results due to damaging the samples by X-rays,[7] we measured the Ce 3d region first. After all other regions were acquired, we remeasured the Ce 3d region again to assess the impact of X-rays on the Au@$CeO_x$ samples. Comparison between $x$ in $CeO_x$ derived from the first (empty data points) and the second Ce 3d measurements (full data points) are plotted in Figure 5S (b) for the samples Ce16.5 and Ce16.0. The shaded areas visualize difference between two measurements. We can see that X-rays indeed induce a $Ce^{4+}$ - to $Ce^{3+}$ reduction of the $CeO_x$ shells. Although X-rays induce the $Ce^{4+}$ - to $Ce^{3+}$ reduction comparable to the reduction caused by laser irradiation at the initial stages of the experiment, the X-rays continuously lose their effect. The reduction by laser irradiation becomes prevalent already after the first laser irradiation.

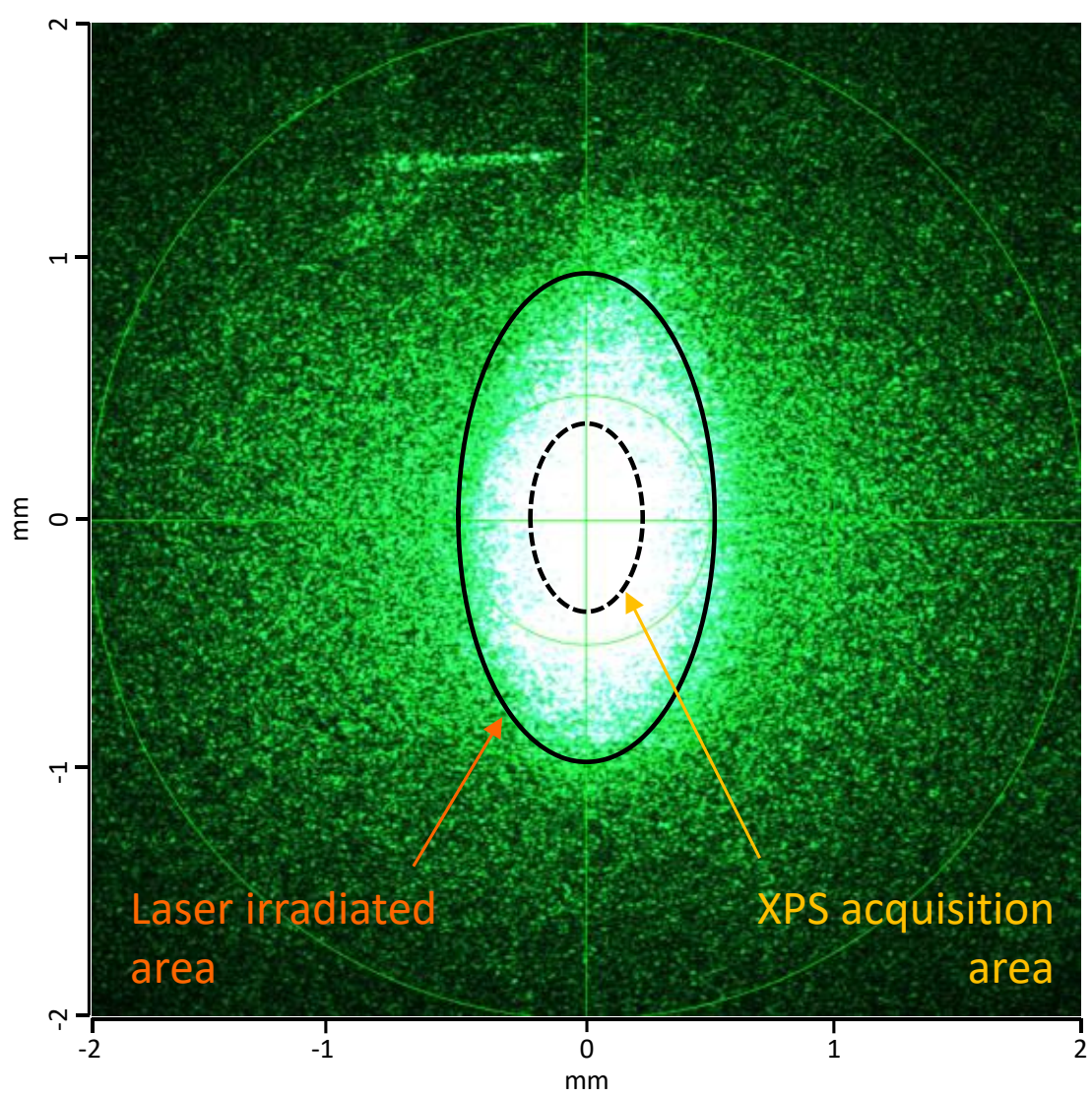

*Figure S4: Image of the sample Ce0.6 taken by an optical microscope in the AXIS Supra photoelectron spectrometer. The area irradiated by the green laser and the area of XPS acquisition are highlighted.*

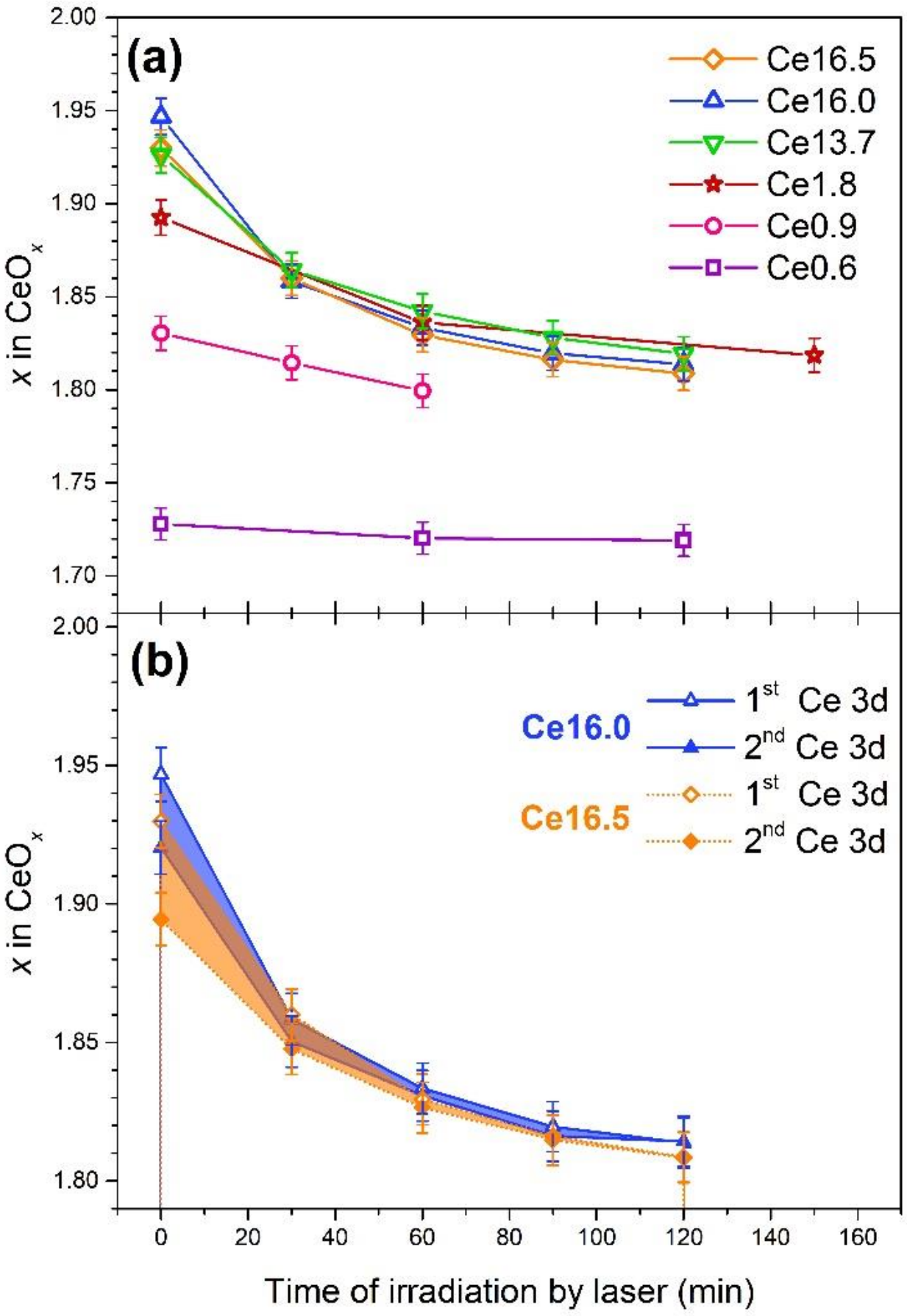

*Figure S5: (a) Evolution of x in $CeO_x$ with prolonged time of laser irradiation. (b) Influence of X-rays on Au@$CeO_x$ core@shell nanoparticles.*

### 3.2 Analysis of XPS data

Binding energies in the photoelectron spectra were calibrated to 916.7 eV of the Ce $3d_{3/2}$ $f^0$ peak. Atomic concentrations of elements were calculated using the ESCApe – Version 1.6.1.1234 software provided by Kratos Analytical (transmission function of the spectrometer and relative sensitivity factors are incorporated in the software and accounted for in the calculations). For purposes of quantitative analysis, the C 1s and Si 2p/Si 2s regions were deconvoluted in the ESCApe software by peaks of gaussian-lorentzian profile with 0.3 lorentzian contribution using a Shirley background.[8–10] An example of the fits is shown in Figure S6. Deconvolution of the Ce 3d region was performed in the KolXPD – Version 1.8.0 (build 68). Atomic concentrations of elements and their bonding states are listed in Table S-III.

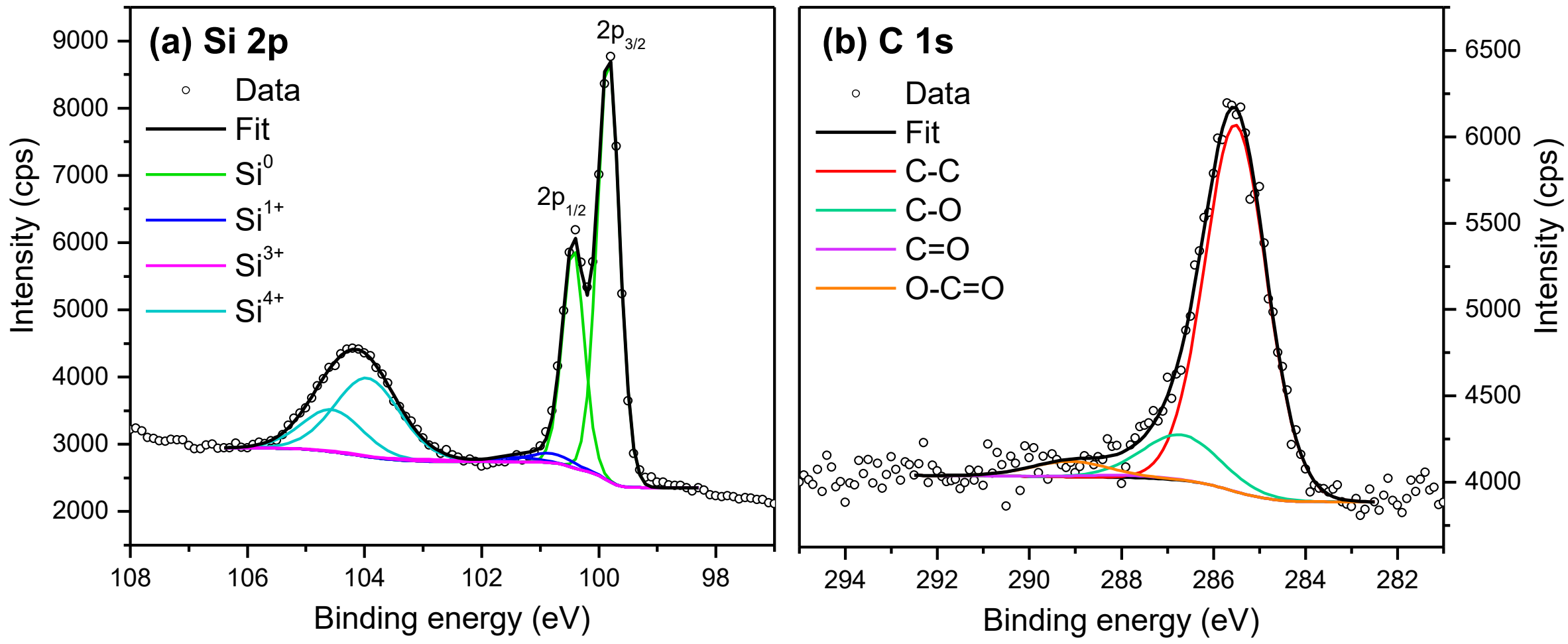

*Figure S6: (a) XPS Si 2p region and (b) C 1s region for the sample Ce0.9. Circles represent measured data, black lines fit results and colorful lines fitting components.*

The following procedure describes evaluation of the amount of oxygen species bound to cerium (including molecular water and -OH species). We estimated the atomic concentrations of oxygen species bound to silicon and carbon and subtracted them from the total concentration of oxygen. We identified C-C, C-O, C=O, and COOH components in the C 1s spectra. In Si 2p or Si 2s, we identified the $Si^{4+}$, $Si^{3+}$, $Si^{+}$ and $Si^{0}$ components. The number of oxygen atoms contributing to one carbon or silicon atom from the C-O, C=O, COOH, $Si^{4+}$, $Si^{3+}$ and $Si^{+}$ bonding states is 1, 1, 2, 2, 1.5 and 0.5, respectively. Therefore, we can estimate the atomic concentration of oxygen species bound to cerium, marked as $O_{Ce}$ and calculated as follows:

$$O_{Ce} = c_O - (1 * c_{C-O} + 1 * c_{C=O} + 2 * c_{COOH} + 2 * c_{Si^{4+}} + 1.5 * c_{Si^{3+}} + 0.5 * c_{Si^{1+}}), \quad \text{(S1)}$$

where *c* refers to an atomic concentration of an oxygen-related non-cerium component. The parameter $O_{Ce}$/Ce in Table S-III refers to the ratio of the $O_{Ce}$ concentration to the total concentration of cerium. Ideally, it should be equal to *x* in $CeO_x$ derived from fitting the Ce 3d spectra. However, this is only true for the initial state of the sample Ce16.0. In all other cases, the values differ strikingly, which is caused mainly by a contribution originating from water species, which adds to the $O_{Ce}$/Ce significantly.

The absolute uncertainty of *x* and $O_{Ce}$/Ce reaches up to 5 % and 30 %, respectively. It is inherently very large but because the procedures during data analysis were done in strictly the same way, the systematic uncertainty is identical for all calculations and significantly reduces the absolute uncertainty. The accuracy of the method to determine *x* in $CeO_x$ used in this study is ± 2 %, and the method is able to detect changes of the $Ce^{3+}$ and $Ce^{4+}$ concentrations between two samples as small as ± 0.5 %.[11] Therefore, we can infer some general trends from parameters $O_{Ce}$/Ce and *x* listed in Table S-III.

*Table S-III: Results of quantitative analysis of XPS data.*

| ¨ | Time (min) | Concentration of elements (at. %) | | | | | Concentration of C- and Si-related oxygen components (at. %) | | | | | | $O_{Ce}/Ce$ | X in $CeO_x$ |
|---|---|---|---|---|---|---|---|---|---|---|---|---|---|---|
| | | Ce | O | C | Au | Si | C-O | C=O | COOH | $Si^{4+}$ | $Si^{3+}$ | $Si^{1+}$ | | |
| Ce16.5 | 0 | 19.3 | 56.0 | 11.7 | 0.5 | 12.4 | 2.7 | 0.0 | 0.0 | 10.7 | 1.7 | 0.0 | 1.52 | 1.93 |
| | 30 | 22.0 | 55.2 | 11.0 | 0.4 | 11.3 | 3.0 | 0.0 | 0.0 | 9.9 | 1.4 | 0.0 | 1.38 | 1.86 |
| | 60 | 21.3 | 54.8 | 10.4 | 0.5 | 13.1 | 2.3 | 0.0 | 0.0 | 11.5 | 1.6 | 0.0 | 1.27 | 1.83 |
| | 90 | 21.0 | 54.8 | 10.7 | 0.5 | 12.9 | 2.3 | 0.0 | 0.0 | 11.4 | 1.5 | 0.0 | 1.31 | 1.82 |
| | 120 | 20.4 | 54.4 | 11.6 | 0.5 | 13.1 | 3.3 | 0.0 | 0.0 | 12.1 | 1.0 | 0.0 | 1.25 | 1.81 |
| Ce16.0 | 0 | 20.0 | 59.8 | 10.8 | 0.6 | 8.8 | 2.5 | 0.7 | 0.2 | 7.7 | 1.1 | 0.0 | 1.96 | 1.95 |
| | 30 | 22.6 | 56.9 | 10.9 | 0.6 | 9.1 | 2.7 | 2.1 | 0.0 | 8.5 | 0.6 | 0.0 | 1.51 | 1.86 |
| | 60 | 19.8 | 55.5 | 13.6 | 0.7 | 10.5 | 2.4 | 3.1 | 0.5 | 9.8 | 0.7 | 0.0 | 1.43 | 1.83 |
| | 90 | 20.1 | 55.8 | 12.9 | 0.7 | 10.5 | 2.1 | 2.9 | 0.0 | 9.5 | 1.0 | 0.0 | 1.51 | 1.82 |
| | 120 | 22.4 | 55.7 | 12.2 | 0.6 | 9.0 | 1.9 | 2.3 | 0.5 | 8.4 | 0.6 | 0.0 | 1.46 | 1.81 |
| Ce13.7 | 0 | 17.0 | 56.1 | 10.6 | 0.9 | 15.4 | 4.1 | 0.4 | 0.3 | 13.5 | 1.9 | 0.0 | 1.24 | 1.93 |
| | 30 | 19.3 | 55.6 | 10.6 | 0.8 | 13.5 | 4.0 | 1.0 | 0.0 | 12.0 | 1.5 | 0.0 | 1.26 | 1.86 |
| | 60 | 20.2 | 54.8 | 10.4 | 0.9 | 13.7 | 4.0 | 0.9 | 0.0 | 12.4 | 1.3 | 0.0 | 1.15 | 1.84 |
| | 90 | 20.3 | 54.5 | 10.2 | 0.9 | 14.2 | 4.0 | 1.4 | 0.0 | 12.8 | 1.4 | 0.0 | 1.05 | 1.83 |
| | 120 | 18.4 | 53.8 | 11.2 | 1.0 | 15.6 | 4.6 | 1.1 | 0.0 | 14.2 | 1.4 | 0.0 | 0.96 | 1.82 |
| Ce8.5 | 0 | 0.7 | 34.1 | 3.7 | 0.1 | 61.4 | 0.7 | 0.1 | 0.1 | 11.4 | 1.4 | 1.3 | 10.79 | 1.93 |
| Ce2.3 | 0 | 9.5 | 45.8 | 12.4 | 14.7 | 17.5 | 0.0 | 0.0 | 0.4 | 9.3 | 2.2 | 2.0 | 2.33 | 1.90 |
| Ce1.8 | 0 | 3.9 | 49.7 | 8.1 | 7.1 | 31.1 | 0.5 | 0.0 | 0.3 | 9.4 | 1.4 | 0.0 | 7.10 | 1.89 |
| | 150 | 4.6 | 47.7 | 9.5 | 7.3 | 30.9 | 0.3 | 0.0 | 0.6 | 9.3 | 1.4 | 0.0 | 5.54 | 1.82 |
| Ce1.4 | 0 | 5.8 | 48.8 | 7.4 | 5.5 | 32.6 | 0.0 | 0.5 | 0.1 | 14.2 | 2.2 | 1.1 | 2.73 | 1.87 |
| Ce1.2 | 0 | 2.9 | 39.6 | 9.1 | 11.7 | 36.8 | 1.0 | 0.6 | 0.3 | 11.2 | 1.6 | 0.9 | 4.19 | 1.89 |
| Ce0.9 | 0 | 1.8 | 33.4 | 20.6 | 10.1 | 34.2 | 2.2 | 0.4 | 0.4 | 11.6 | 0.3 | 0.9 | 3.28 | 1.83 |
| | 30 | 1.9 | 33.2 | 20.6 | 10.6 | 33.7 | 3.2 | 0.0 | 0.9 | 11.1 | 0.3 | 0.8 | 2.71 | 1.81 |
| | 60 | 2.0 | 33.0 | 19.5 | 10.9 | 34.7 | 2.7 | 0.0 | 0.9 | 11.7 | 0.0 | 0.9 | 2.33 | 1.80 |
| | Air | 1.8 | 35.1 | 18.7 | 11.0 | 33.3 | 3.1 | 0.0 | 1.1 | 11.2 | 0.0 | 1.0 | 3.83 | 1.85 |
| | 60 | 1.9 | 34.6 | 18.6 | 11.3 | 33.5 | 3.4 | 0.0 | 0.6 | 11.1 | 0.0 | 1.1 | 3.82 | 1.81 |
| | UHV | 1.9 | 34.2 | 19.6 | 11.3 | 33.1 | 2.5 | 0.1 | 0.9 | 11.4 | 0.0 | 0.7 | 3.50 | 1.81 |
| Ce0.6 | 0 | 0.1 | 68.2 | 7.3 | 0.5 | 23.9 | 1.9 | 0.0 | 0.4 | 22.3 | 1.1 | 0.0 | 192.50 | 1.73 |
| | 60 | 0.1 | 63.7 | 12.6 | 0.5 | 23.2 | 3.3 | 0.3 | 0.4 | 21.3 | 1.5 | 0.0 | 144.50 | 1.72 |
| | 120 | 0.1 | 61.7 | 15.4 | 0.5 | 22.2 | 3.6 | 0.3 | 0.1 | 20.8 | 1.1 | 0.0 | 143.50 | 1.72 |